\documentclass[conference]{IEEEtran}
\IEEEoverridecommandlockouts
\usepackage{amsmath}
\usepackage{amsfonts}
\usepackage{amssymb}
\usepackage{mathtools}
\usepackage{amsthm}
\usepackage{fontenc}
\usepackage{graphicx}
\usepackage[linesnumbered,ruled]{algorithm2e}
\usepackage{booktabs}
\usepackage{subfigure}
\usepackage{tabulary}
\usepackage{booktabs}
\usepackage[justification=centering]{caption}
\usepackage{comment}

\usepackage{cite}

\newtheorem{lemma}{\textbf{Lemma}}

\SetKw{KwBy}{by}
\usepackage[colorinlistoftodos]{todonotes}
\usepackage{caption}
\newcommand{\mv}[1]
{\mbox{\boldmath{$#1$}}}

\title{Zero-shot Multi-level Feature Transmission Policy  Powered by Semantic Knowledge Base}
\author{Yaping Sun, Hao Chen, Xiaodong Xu,  Ping Zhang,~\IEEEmembership{Fellow,~IEEE}, Shuguang Cui,~\IEEEmembership{Fellow,~IEEE}

\thanks{Y.~Sun and H.~Chen are with the Department of Broadband Communication, Peng Cheng Laboratory, Shenzhen 518000, China. (email: \{sunyp, chenh03\}@pcl.ac.cn)}
\thanks{X.~Xu and P.~Zhang are  with the Beijing University of Posts and Telecommunications, Beijing 100876, China, and affiliated with the Department of Broadband Communication, Peng Cheng Laboratory, Shenzhen 518000, China. (email: xuxiaodong, pzhang@bupt.edu.cn)}
\thanks{S.~Cui is with the School of Science and Engineering (SSE) and \textcolor{black}{the Future Network of Intelligent Institute (FNii)}, the Chinese University of Hong Kong (Shenzhen), Shenzhen 518172, China. S.~Cui is also with Shenzhen Research Institute of Big Data, Shenzhen 518172, China, and affiliated with the Department of Broadband Communication, Peng Cheng Laboratory, Shenzhen 518000, China (email: shuguangcui@cuhk.edu.cn).}
}

\begin{document}
\maketitle

\begin{abstract}
Remote zero-shot object recognition, i.e., offloading zero-shot object recognition task from one mobile device to remote mobile edge computing (MEC) server or another mobile device, has become a common and important task to solve for 6G. In order to tackle this problem, this paper first establishes a zero-shot multi-level feature extractor, which projects the image into visual, semantic, as well as intermediate feature space in a lightweight way. Then, this paper  proposes a novel multi-level feature transmission framework powered by semantic knowledge base (SKB), and characterizes the semantic loss and required transmission latency at each level. Under this setup, this paper formulates the multi-level feature transmission optimization problem to minimize the  semantic loss under the end-to-end latency constraint.  The optimization problem, however, is a multi-choice knapsack problem, and thus very difficult to be optimized. To resolve this issue, this paper proposes an efficient algorithm based on convex concave procedure to find a high-quality solution. Numerical results show that the proposed  design outperforms the benchmarks, and illustrate the tradeoff between the transmission latency and zero-shot classification accuracy, as well as the effects of  the SKBs at both the transmitter and receiver on classification accuracy. 
\end{abstract}


\section{Introduction}

Recent technical advancements of wireless communication and artificial intelligence (AI) have enabled multiple emerging applications, e.g., auto-driving, mixed reality, metaverse, industrial internet, etc. Due to the diversity and time variety of the data distributions in such scenarios, \textbf{remote zero-shot object recognition/learning} has become one of the most common and important problems to solve. For example,  in order to support automatic navigation or avoid collision, vehicles have to share/receive and process traffic 
information in real time to/from the remote vehicles 
or the road side units (RSUs) based on classifiers and wireless communication to recognize the remote traffic situation \cite{userass}. However, there are often new traffic categories such as emergency accidents and new vehicles, which are not available in advance and cannot be seen during training of the classifier models. Such tasks for vehicles can be named as remote zero-shot object recognition. 

The challenges for solving the remote zero-shot object recognition problem mainly lie in two aspects. The first is \textit{the zero-shot object recognition problem}, i.e., recognizing novel image categories without any training samples. Different from traditional supervised deep learning (DL)-based classifiers, which generally require hundreds or thousands of training samples and also retraining the DL models to recognize a new category, achieving zero-shot recognition would help significantly reduce computation,  communication, caching, as well as time consumptions, and is in line with the ``intellicise" development vision of future 6G \cite{intellicise}. 

The second is \textit{the mobile remote recognition problem}, i.e., offloading the recognition task to remote mobile edge computing (MEC) server or another mobile device, which involves the sensing-preprocessing-communicating-postprocessing-recognition service loop. To tackle the remote recognition problem, existing literature mainly consists of two approaches, i.e.,  edge inference and task-oriented semantic communication. 
Edge inference, i.e., conducting the inference task at the edge of wireless networks such as mobile devices and MEC  servers, can eliminate the transmission and routing latency from the edge to the cloud, and reduce the service latency and communication bandwidth requirement. In this line of research, three different types of edge inference approaches have been considered, namely device-only inference \cite{device1}, edge-only inference \cite{edgeonly}, as well as device-edge co-inference \cite{coinfersingle,coinfermulti,coinfer1}. However, the existing literature on edge inference has not tackled the zero-shot object recognition problem. 

Another promising approach is task-oriented semantic communication, whereby transmitters are designed to efficiently convey semantic information relevant to the tasks to receivers, rather than reliably transmit syntactic information as in conventional wireless communication systems \cite{6G3}. Via the end-to-end (E2E) joint semantic-channel coding/decoding design, the semantic communications are able to efficiently compress messages while preserving the essential meaning by filtering out the task-irrelevant information, and thus significantly enhance the communication efficiency \cite{tasksem1, xietask, masked}. However, the aforementioned works \cite{tasksem1,xietask, masked} rely on large-scale labled training datasets, and have not considered the zero-shot recognition problem yet. In addition, the joint source-channel coding framework contradicts the conventional separate coding module, and thus cannot be directly applied for the  existing communication networks. 


Thus, solving the E2E remote zero-shot recognition problem is of great importance and challenges for the realization of future 6G intellicise network.  Similar to humans' knowledge-based recognition, i.e., humans can transfer their knowledge to identify new classes when only textual descriptions of the new classes are available, this paper considers multi-level feature transmission powered by semantic knowledge base (SKB) to support remote zero-shot recognition. The main contributions of this paper are listed as below.
\begin{itemize}
    \item First, to support the zero-shot learning task and get rid of dependence on big datasets, a lightweight multi-level feature extractor powered by SKB  is proposed, which consists of intermediate feature extractor, visual autoencoder as well as semantic autoencoder. 
    \item Then, based on the aforementioned multi-level feature extractor, a multi-level feature transmission model powered by SKB is established, in which both transmitter and receiver are enabled with SKB and multi-level feature extractor. Then, the semantic loss minimization problem under the transmission latency constraint is formulated, which is a multi-choice knapsack problem and is generally NP-hard \cite{multichoice}. In order to reduce the computation complexity, convex concave procedure (CCCP) method is adopted to achieve an efficient sub-optimum.   
    \item Finally, numerical results show the promising performance gains of the proposed design, as compared with conventional designs without such multi-level optimization. The proposed multi-level transmission designs are observed to better utilize the knowledge at both the transmitter and receiver to achieve promising zero-shot classification under the transmission latency constraint. 
\end{itemize}





%

 \section{SKB-enabled Multi-level Feature Extractor} 

Consider $N$ training image samples $\mathcal{N}$, represented with $\left(\mv V, \mv C,\mv S\right)$. In particular, $\mv V \in \mathbb{R}^{D_v\times N}$ denotes the visual feature matrix of the $N$ image samples, which is extracted by pretrained deep convolutional neural networks (CNNs). For example,  the visual features can be GoogleNet features, which are the $1024$-dimensional activations of the final pooling layer as in \cite{pretrainfeature}. $\mv C \triangleq (c_n)_{n\in \mathcal{N}} \in \mathcal{C}_{\text{Tr}}^{N}$ denotes the class label vector of the image samples, where $c_n \in \mathcal{C}_{\text{Tr}}$ denotes the class label of sample $n$, and $\mathcal{C}_{\text{Tr}}$ denotes the class set seen within the training samples. $\mv S \in \mathbb{R}^{D_s\times N}$ denotes the semantic feature matrix of the image samples, each column of which corresponds to the semantic feature vector of the class $c_n$.

\subsection{Intermediate feature extractor}
First, a low-dimensional intermediate feature extractor is designed via extending the conditional principal label space transformation (CPLST) approach \cite{hiddenvec}. The benefits of this novel approach lie in two aspects. On one hand, it allows to reduce the feature space into a lower and controllable dimension. On the other hand, it considers both  visual and semantic feature, and thus can bridge the gap between the statistical property of the visual features and that of the semantic features. 

\subsubsection{Formulation}
Specifically, the visual feature $\mv V$ and semantic feature $\mv S$ are projected into a $k$-dimensional latent space with a visual projection matrix $\mv W_v \in \mathbb{R}^{k\times D_v}$ and a semantic projection matrix $\mv W_s \in \mathbb{R}^{k \times D_s}$, respectively, where $k\leq \min \{D_v,D_s\}$.  Similar to \cite{hiddenvec}, the predicting error and encoding error are minimized simultaneoursly:
\begin{align}
\text{(P1)}\ \ \  & \min_{\mv W_v, \mv W_s} \ \ \   \Arrowvert \mv W_v \mv V - \mv W_s \mv S \Arrowvert_F^2 + \Arrowvert \mv S - \mv W_s^T \mv W_s \mv S \Arrowvert
 ^2_F\nonumber\\ 
  &\ \ \ \  s.t. \ \ \ \ \ \ \ \ \ \mv W_s \mv W_s^T = \mv I. \nonumber
\end{align} 
Compared with the traditional CPLST approach,  the semantic feature instead of binary label feature is utilized to characterize the semantic  relationship  among classes.

\subsubsection{Optimization}
First, given $\mv W_s$, the closed-form optimal $\mv W_v^*$, i.e., $\mv W_v^* = \mv W_s \mv S \mv V^{\dag}$ is directly obtained, where $\mv V^{\dag}$ is the pesudo inverse of $\mv V$. Then, via substituting $\mv W_v$ with $\mv W_s \mv S \mv V^{\dag}$, problem~(P1) is equally transformed into
\begin{align}
 & \max_{\mv W_s} \ \ \   \text{tr}\left(\mv W_s \mv S \mv H \mv S^T \mv W_s^T \right)\nonumber\\ 
  &\  s.t. \ \ \ \ \ \ \ \ \ \mv W_s \mv W_s^T = \mv I,\nonumber
\end{align} 
where $\mv H = \mv V^{\dag} \mv V$. 

According to Eckart-Young theorem \cite{EYtheorem}, the optimal solution is given by  the eigenvectors that correspond to the largest eigenvalues of $\mv S \mv H \mv S^T$. 

\subsubsection{Common intermediate feature}
After getting $\mv W_s$, the semantic feature $\mv S$ is linearly mapped to the intermediate feature vector $\mv F$ by $\mv F = \mv W_s \mv S$, which is used in the sequel.

\subsection{Visual autoencoder}
\subsubsection{Formulation}
Different from  the conventional autoencoder which is unsupervised, the latent space is forced to be the low-dimensional intermediate feature $\mv F$. The learning objective is transformed into 
\begin{align}
(\text{P2})\ \ \  & \min_{\mv P_v} \ \ \  \Arrowvert \mv V - \mv P_v^T \mv P_v \mv V \Arrowvert
 ^2_F \nonumber\\ 
  &\  s.t. \ \ \ \ \ \ \ \ \ \mv P_v \mv V = \mv F,\nonumber
\end{align} 
where $\mv P_v \in \mathbb{R}^{k\times D_v}$ denotes the visual projection matrix which maps the visual feature $\mv V$ into the intermediate  feature $\mv F$. 
\subsubsection{Optimization}
Similar to \cite{SeAE}, to optimize problem~(P2), the strict equal constraint is firstly relaxed into its objective, i.e.,
\begin{align}
  (\text{P3}) \ \  \min_{\mv P_v} \ \ \  \Arrowvert \mv V - \mv P_v^T \mv P_v \mv V \Arrowvert^2_F + \lambda \Arrowvert \mv P_v\mv V - \mv F \Arrowvert^2_F, \nonumber
\end{align}
where $\lambda$ is a weight factor which controls the tradeoff between the loss of the encoder, i.e., the first item, and that of the decoder, i.e., the second item. 

Then, the optimal solution to problem~(P3) is obtained via setting the first-order derivative of its objective to zero, i.e.,
\begin{align}
    \mv F\mv F^T \mv P_v + \lambda \mv P_v  \mv V \mv V^T  = (1+\lambda)\mv F \mv V^T,  
\end{align}
which is the well-known Sylvester equation, and can be optimally solved via the Bartels-Stewart algorithm \cite{SeAE}. 

\subsection{Semantic autoencoder}
\subsubsection{Formulation}
Similar to the visual autoencoder,  a semantic autoencoder is also designed which forces the semantic feature to be projected into the intermediate feature $\mv F$. In particular, the learning objective is
\begin{align}
(\text{P4})\ \ \  & \min_{\mv P_s} \ \ \  \Arrowvert \mv S - \mv P_s^T \mv P_s \mv S \Arrowvert
 ^2_F \nonumber\\ 
  &\  s.t. \ \ \ \ \ \ \ \ \ \mv P_s \mv S = \mv F,\nonumber
\end{align} 
where $\mv P_s \in \mathbb{R}^{k\times D_s}$ denotes the semantic projection matrix which maps the semantic feature $\mv S$ into the intermediate  feature $\mv F$. 
\subsubsection{Optimization}
The optimal solution to (\text{P4}) can be obtained via the same way of solving (P2). 
First, the strict equal constraint is relaxed into its objective, i.e.,
\begin{align}
  (\text{P5}) \ \  \min_{\mv P_s} \ \ \  \Arrowvert \mv S - \mv P_s^T \mv P_s \mv S \Arrowvert^2_F + \lambda \Arrowvert \mv P_s\mv S - \mv F \Arrowvert^2_F, \nonumber
\end{align}
where $\lambda$ is a weight factor which controls the tradeoff between the loss of the encoder, i.e., the first item, and that of the decoder, i.e., the second item. 

Then, the optimal solution to problem~(P5) is obtained via setting the first-order derivative of its objective to zero, i.e.,
\begin{align}
    \mv F\mv F^T \mv P_s + \lambda \mv P_s  \mv S \mv S^T  = (1+\lambda)\mv F \mv S^T,  
\end{align}
which is the well-known Sylvester equation, and can be optimally solved via the Bartels-Stewart algorithm \cite{SeAE}.
\subsection{Multi-level feature extractor}
Based on the above-mentioned visual and semantic autoencoder, given any image sample $\mv l$, we design the following multi-level feature extractor:
\begin{itemize}
    \item $1$st-level visual feature, i.e., $\mv v = f(\mv l) \in \mathbb{R}^{D_v}$, which  is obtained via projecting the image sample $\mv l$ with the pre-trained large scale CNNs  $f(\cdot)$. 
    \item $2$nd-level intermediate feature, i.e.,  $\mv f = \mv P_v \mv v \in \mathbb{R}^k$, which is obtained via projecting the visual feature $\mv v$ with the visual encoder $\mv P_v$.
    \item $3$rd-level semantic feature, i.e.,  $\mv s = \mv P_s^T \mv f \in \mathbb{R}^{D_s}$, which is obtained via projecting the intermediate feature $\mv f$ with the semantic decoder $\mv P_s^T$.
    \item $4$th-level estimated class label, i.e., $\hat{c} = \arg \min_{c\in \mathcal{C}} \Arrowvert \mv s_c - \mv s \Arrowvert^2_F$, where $\mathcal{C}$ denotes the available class set, and $\mv s_c$ denotes the semantic vector of class $c$. 
\end{itemize}


     \begin{figure}[t]
    \centering
    \includegraphics[width=.5\textwidth]{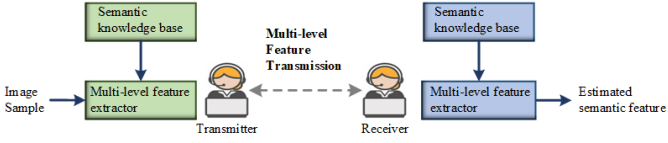}
   \caption{ SKB-enabled  E2E communication system.} \label{fig:e2e}
\end{figure}

\section{System Model}

As illustrated in Fig.~\ref{fig:e2e}, a novel E2E communication system is considered, where both transmitter and receiver are enabled with a specific semantic knowledge base (SKB) and multi-level feature extractor. 

\subsection{Semantic knowledge base (SKB)}
We define SKB of each mobile device as a   set of semantic vectors of some classes stored at it. Based on local SKB, the mobile device is able to recognize  the attributes of corresponding classes and the semantic relationship among the classes. Specifically, denote with $\mathcal{C} \triangleq \left\{1,2,\cdots,C\right\}$ the set of all classes, and $\mathcal{S} \triangleq \left\{\mv s_1, \mv s_2, \cdots, \mv s_C\right\}$ the set of semantic vectors of all the classes, named as semantic prototype. The SKBs at both transmitter and receiver are modeled as follows:
\begin{itemize}
    \item SKB at the transmitter: let $t_c \in \{0,1\}$ denote the semantic knowledge indicator of class $c$ at the transmitter, where $t_c = 1$ indicates that the transmitter has the knowledge of semantic information of class $c$, i.e., the semantic vector $\mv s_c$ of class $c$ is stored at the transmitter, and $t_c = 0$, otherwise.  Denote with $\mathcal{B}_T \triangleq \{c\in \mathcal{C}: t_c = 1\}$ the set of class labels, the semantic vectors of which are stored at the transmitter, i.e., the SKB at the transmitter. 
    \item SKB at the receiver: let $r_c \in \{0,1\}$ denote the semantic knowledge indicator of class $c$ at the receiver, where $r_c = 1$ indicates that the receiver has the knowledge of semantic information of class $c$, i.e., the semantic vector $\mv s_c$ of class $c$ is stored at the receiver, and $r_c = 0$, otherwise.  Denote with $\mathcal{B}_R \triangleq \{c\in \mathcal{C}: r_c = 1\}$ the set of class labels, the semantic vectors of which are stored at the receiver, i.e., the SKB at the receiver. 
\end{itemize}

\subsection{Multi-level feature transmission policy}

Suppose that both transmitter and receiver are enabled with its own multi-level feature extractor, which is trained from its own training dataset, denoted as $(\mv V_t, \mv C_t, \mv S_t)$ and $(\mv V_r, \mv C_r, \mv S_r)$, respectively. Let $\mv P_{t,v}$ and $\mv P_{t,s}$ ($\mv P_{t,v}^T$ and $\mv P_{t,s}^T$) denote the visual and semantic encoder (decoder) at the transmitter, respectively. And let $\mv P_{r,v}$ and $\mv P_{r,s}$ ($\mv P_{r,v}^T$ and $\mv P_{r,s}^T$) denote the visual and semantic encoder (decoder) at the receiver, respectively. 

Consider that there are $M$ testing samples, denoted as $\mathcal{M} \triangleq \{1,2,\cdots,M\}$, which have not been seen during the training process at both the transmitter and receiver, and require to be classified. Based on the above-mentioned multi-level feature extractor, there are the following four kinds of transmission choice at the transmitter for each testing sample to complete the zero-shot classification task. Let $x_{m,l} \in \{0,1\}$ denote the transmission choice of sample~$m$, where $x_{m,l}=1$ indicates that the $l$-th level feature vector of sample $m$ is transmitted, and $x_{m,l} = 0$, otherwise. In order to guarantee that the semantic information of each sample is delivered to the receiver, we have $\sum_{l=1}^4 x_{m,l} = 1, \forall m \in \mathcal{M}$. 

\begin{itemize}
    \item $1$st-level visual feature transmission: when $x_{m,1} =1$, the visual feature vector of sample $m$, denoted as $\mv v_m \in \mathbb{R}^{D_v}$, is directly transmitted to the receiver. Then, the receiver estimates the class based on its multi-level feature extractor and its SKB, i.e., $\hat{c}_{r,m} = \arg \min_{c\in \mathcal{B}_R} \Arrowvert \mv s_c - \mv P_{r,s}^T \mv P_{r,v}\mv v_m \Arrowvert^2_F$. And the corresponding semantic loss is given by $ L_{m,1} = \min_{c\in \mathcal{B}_R} \Arrowvert \mv s_{c} - \mv P_{r,s}^T \mv P_{r,v}\mv v_m \Arrowvert^2_F$. The required transmission latency is given by $T_{m,1} = \frac{D_vQ}{R}$, where $Q$ denotes the quantization level \cite{coinfersingle} and $R$ denotes the achievable data rate from the  transmitter to the receiver.\footnote{For ease of analysis, uniform quantization of each vector is adopted for digital transmission as in \cite{coinfermulti} such that each element of the vector is quantized into an equal number of bits throughout this paper.}
    \item $2$nd-level intermediate feature transmission: when $x_{m,2}=1$, the intermediate feature vector of sample $m$, i.e., $\mv f_m = \mv P_{t,v}\mv v_m \in \mathbb{R}^{k}$, is transmitted to the receiver. Then, the receiver estimates the class based on its semantic decoder, i.e., $\hat{c}_{r,m} = \arg \min_{c \in \mathcal{B}_R} \Arrowvert \mv s_c - \mv P_{r,s}^T \mv f_m \Arrowvert^2_F$. And the corresponding semantic loss  is given by $ L_{m,2} = \min_{c\in \mathcal{B}_R} \Arrowvert \mv s_{c} - \mv P_{r,s}^T \mv P_{t,v}\mv v_m \Arrowvert^2_F$. The required transmission latency is given by $T_{m,2} = \frac{kQ}{R}$. 
    \item $3$rd-level semantic feature transmission: when $x_{m,3}=1$, the semantic feature vector of sample $m$, i.e., $\mv s_m = \mv P_{t,s}^T\mv P_{t,v}\mv v_m \in \mathbb{R}^{D_s}$, is transmitted to the receiver. Then, the receiver estimates the class based on its SKB, i.e., $\hat{c}_{r,m} = \arg \min_{c\in \mathcal{B}_R} \Arrowvert \mv s_c - \mv s_m \Arrowvert^2_F$. And the corresponding semantic loss  is given by $ L_{m,3} = \min_{c\in \mathcal{B}_R} \Arrowvert \mv s_{c} - \mv P_{t,s}^T \mv P_{t,v}\mv v_m \Arrowvert^2_F$. The required transmission latency is given by $T_{m,3} = \frac{D_sQ}{R}$.
    \item $4$th-level estimated class knowledge transmission: when $x_{m,4}=1$, the transmitter first estimates the class of the image sample, i.e., $\hat{c}_{t,m} = \arg \min_{c\in \mathcal{B}_T} \Arrowvert \mv s_c - \mv P_{t,s}^T\mv P_{t,v} \mv v_m \Arrowvert^2_F$, and then transmits the estimated label $\hat{c}_{t,m}$ or the corresponding semantic vector $\mv s_{\hat{c}_{t,m}}$ to the receiver, according to the SKB at the receiver. Specifically, if the semantic vector of class $\hat{c}_{t,m}$ is stored in the SKB of the receiver, i.e., $r_{\hat{c}_{t,m}} = 1$, transmitting the estimated label is sufficient, and the transmission load is $Q$. Otherwise, i.e., $r_{\hat{c}_{t,m}} = 0$, the semantic vector of the estimated class $\mv s_{\hat{c}_{t,m}}$ requires to be transmitted to the receiver, and the transmission load is $D_sQ$. Thus, the required transmission latency is given by $T_{m,4} = \frac{\left[r_{\hat{c}_{t,m}} + D_s\left(1-r_{\hat{c}_{t,m}}\right)\right]Q}{R}$.
    And the corresponding semantic loss is given by $L_{m,4} = \min_{c\in \mathcal{B}_T} \Arrowvert  \mv s_{c} - \mv P_{t,s}^T \mv P_{t,v}\mv v_m \Arrowvert^2_F$. 
\end{itemize}

In summary, the average semantic loss deemed by the E2E communication is given by
\begin{align}
    \frac{1}{M}\sum_{m=1}^M \sum_{l=1}^4 x_{m,l} L_{m,l},
\end{align}
and the average transmission latency constraint is given by
\begin{align}
   \frac{1}{M} \sum_{m=1}^M \sum_{l=1}^4 x_{m,l} T_{m,l} \leq \tau.
\end{align}

\section{Problem Formulation and Convex Concave Procedure}
\subsection{Problem formulation}
Under this setup, our objective is to minimize the semantic loss based on the knowledge of both transmitter and receiver via optimizing the multi-level feature transmission policy $\mv x \triangleq \left(x_{m,l}\right)_{m\in \mathcal{M},l\in \{1,2,3,4\}}$. The optimization problem is formulated as
\begin{align}
    (\text{P6})\ \ \ & \min_{\mv x}\ \ \ \ \  \sum_{m=1}^M\sum_{l=1}^4 x_{m,l}L_{m,l} \nonumber  \\
    &\ s.t. \ \ \ \frac{1}{M} \sum_{m=1}^M \sum_{l=1}^4 x_{m,l}T_{m,l}\leq \tau, \label{eq:taucons}\\
    &\ \ \ \ \ \ \ \ \sum_{l=1}^4 x_{m,l} = 1, \forall m \in \mathcal{M},\label{eq:sum1}\\
    &\ \ \ \ \ \ \ \ x_{m,l} \in \{0,1\}, \forall m \in \mathcal{M}, l \in \{1,2,3,4\}.\label{eq:binary}
\end{align}

It can be observed that problem~(P6) is a linear multi-choice knapsack problem, which is NP-hard \cite{multichoice}.\footnote{For ease of feasibility of problem~(P6), it is assumed that $\frac{Q}{R} \leq \tau$ throughout this paper.} Denote with $\mv x^* \triangleq (x^*_{m,l})_{m\in \mathcal{M},l \in \{1,2,3,4\}}$ the optimal multi-level transmission policy of problem~(P6). 
Notice that there exists a tradeoff between the semantic loss and transmission latency, and  where to extract the feature (i.e., whether to utilize the multi-level feature extractor at the transmitter or that at the receiver), which level to extract, as well as where to make the semantic information inference decision (i.e., whether to utilize the SKB at the transmitter or that at the receiver) have to be carefully designed to minimize the semantic information loss, while guaranteeing the transmission latency constraint.

\subsection{Convex concave procedure (CCCP)}
In this section,  problem~(P6) is solved via CCCP. First, constraint (\ref{eq:binary}) is rewritten as
\begin{align}
   & x_{m,l} \in \left[0,1\right], \forall m \in \mathcal{M}, l \in \{1,2,3,4\},\label{eq:eqc1}\\
   & x_{m,l}\left(1-x_{m,l}\right) \leq 0, \forall m \in \mathcal{M}, l \in \{1,2,3,4\}, \label{eq:eqc2}
\end{align}
without loss of equivalence. Then, via substituting constraint (\ref{eq:binary}) with (\ref{eq:eqc1}) and (\ref{eq:eqc2}), problem~(P6) is  equivalently transformed into problem~(P7).
\begin{align}
    (\text{P7})\ \ \ & \min_{\mv x}\ \ \ \ \  \sum_{m=1}^M\sum_{l=1}^4 x_{m,l}L_{m,l} \nonumber  \\
    &\ s.t. \ \ \ \ \ \ (\ref{eq:taucons})(\ref{eq:sum1})(\ref{eq:eqc1})(\ref{eq:eqc2}). \nonumber
\end{align}
Notice that problem~(P7) is a continuous optimization problem,
and thus the computation complexity of solving it is much less than that of solving problem~(P6) directly. However, since constraint (\ref{eq:eqc2}) is a concave constraint, problem~(P7) is a non-convex optimization problem and thus optimizing  problem~(P7) is still very difficult.

Next, to facilitate solving problem~(P7), problem~(P7) is transformed into
problem (P8) by penalizing the concave constraint  (\ref{eq:eqc2}) into the
objective of problem~(P7).
\begin{align}
    (\text{P8})\ \ \ & \min_{\mv x}\ \ \ \ \  \sum_{m=1}^M\sum_{l=1}^4 \left(x_{m,l}L_{m,l}-\gamma x_{m,l}\left(x_{m,l}-1\right) \right)\nonumber  \\
    &\ s.t. \ \ \ \ \ \ (\ref{eq:taucons})(\ref{eq:sum1})(\ref{eq:eqc1}),\nonumber
\end{align}
where $\gamma > 0$ denotes the penalty parameter. Let $\bar{L}(\gamma)$ denote the optimal objective value of problem~(P8). 
\addtolength{\topmargin}{0.03in}

Note that problem (P8) is  an indefinite quadratic programming (IQP) due to its objective function being a difference between a linear function and a quadratic convex function, while its constraints are linear \cite{iqp}. This makes it a special case of the difference of convex problem. By utilizing difference of convex algorithms (DCA), local optima for problem (P8) can be obtained in a finite number of steps. It is worth noting that DCA is exactly the same as CCCP when the second term of the objective function of problem (P8) is differentiable. To solve problem (P8) using CCCP, a sequence of linear optimization problems needs to be solved iteratively, which are obtained by linearizing the second term of the IQP objective function. Specifically, at each
iteration $t$, $ \sum_{m=1}^M\sum_{l=1}^4 x_{m,l}\left(x_{m,l}-1\right)$ is approximated as $ \sum_{m=1}^M\sum_{l=1}^4 x^{(t)}_{m,l}\left(x^{(t)}_{m,l}-1\right) +  \sum_{m=1}^M\sum_{l=1}^4 \left(2x_{m,l}^{(t)}-1\right)\left(x_{m,l}-x_{m,l}^{(t)}\right)$.

In the end,  the equivalence between problem~(P7) and problem~(P8) is demonstrated in Lemma~\ref{lem:exact}. 
\begin{lemma}\label{lem:exact}
For all $\gamma > \gamma_0$ where
\begin{equation}
    \gamma_0 \triangleq \frac{\sum_{m=1}^M\sum_{l=1}^4 x_{m,l}^0L_{m,l} - \bar{L}(0)}{\max_{\mv x} \left\{\sum_{m=1}^M\sum_{l=1}^4 x_{m,l}\left(x_{m,l}-1\right):(\ref{eq:taucons})(\ref{eq:sum1})(\ref{eq:eqc1}) \right\}},\nonumber
\end{equation}
with any $x_{m,l}^0$ satisfying (\ref{eq:taucons}), (\ref{eq:sum1}), and (\ref{eq:eqc1}), problem~(P8) and problem~(P7) have the same optimal solution. 
\end{lemma}

Lemma~\ref{lem:exact} demonstrates that when the penalty parameter $\gamma$ is sufficiently large, problem~(P8) becomes equivalent to problem~(P7). Therefore, solving problem~(P8) using CCCP can replace solving problem~(P7). However, problem~(P7) may not always have a feasible solution. To obtain the global optimum of problem~(P7), CCCP can be performed multiple times, each time with a distinct initial feasible point of problem~(P8), and then the solution which achieves the lowest average value across all runs is chosen \cite{cccpexact}.




\section{Numerical Results}
\begin{table}[t]
\caption{Transmission latency vs. classification accuracy under differnet policies for AWA dataset.}
\begin{center}
\newcommand{\tabincell}[2]{\begin{tabular}{@{}#1@{}}#2\end{tabular}}
\newcommand{\tl}[1]{\multicolumn{1}{l}{#1}} 
\renewcommand\arraystretch{1}
\setlength{\tabcolsep}{2mm}{
\begin{tabular}{|c|c|c|}
\hline
Transmission Policy  &   Transmission Latency & Classification Accuracy \\
\hline
Level~$1$ transmission &  $1074.4$~ms & $67.8\%$\\
\hline
Level~$2$ transmission &  $ 15.7$~ms & $67.8\%$\\
\hline
Level~$3$ transmission & $ 89.2$~ms & $67.8\%$ \\
\hline
Level~$4$ transmission & $35.3$~ms & $68.6\%$\\
\hline
CCCP method & $7.9$~ms & $67.8\%$\\
\hline
\end{tabular}}
\end{center}
\label{tab:notation3}
\end{table}
This section provides numerical results to validate the
performance of the proposed framework and transmission policy. For comparison, the following three kinds of baselines are considered.
\begin{itemize}
    \item Level-$j$ transmission, $j\in \{1,2,3,4\}$: $x_{m,j} = 1$, and $x_{m,j'} = 0$,  $\forall j' \in \{1,2,3,4\}\setminus j$, $ m \in \mathcal{M}$. 
    \item Linear relaxation method:  the binary constraint $x_{m,l} \in \{0,1\}$ is first relaxed into the real constraint $x_{m,l} \in [0,1]$, and then problem~(P6) is transformed into a linear program (LP), which can be optimally solved via standard methods, e.g., CVX. Let $\hat{\mv x} \triangleq (\hat{x}_{m,l})_{m\in \mathcal{M},l\in \{1,2,3,4\}}$ denote the optimal solution to LP. Based on $\hat{\mv x}$, the linear relaxation-based association $\mv x$ is chosen as $x_{m,l^*} = 1$, where $l^* \triangleq \arg \max_{l\in {1,2,3,4}} \hat{x}_{m,l}$, and $x_{m,l} = 0$, otherwise.
    \item Lagrangian relaxation method: 
Another suboptimal solution to problem~(P6) is obtained via Lagrangian relaxation (LR) method  \cite{lr}. 
\end{itemize}

The dataset Animals with Attributes (AwA)  is adopted \cite{awa}. The path loss $g$ between the transmitter and receiver is modeled as $\beta_0 \left(\frac{d}{d_0}\right)^{-\zeta}$, where $\beta_0 = -30$ dB denotes the path loss at the reference distance $d_0 = 10$~m, 
$d = 500$~m denotes the distance between them, and $\zeta = 3$ denotes the path loss exponent. Furthermore, the transmission rate $R$ is set as $B \log_2 \left(1+ \frac{p g}{BN_0}\right)$, where $B = 1$ MHz, $N_0 = -174$ dBm/H, and $p = 10$ dBm.

\begin{figure}[t]
  \centering
  \subfigure[SKB at the transmitter.] {\includegraphics[width=.3\textwidth]{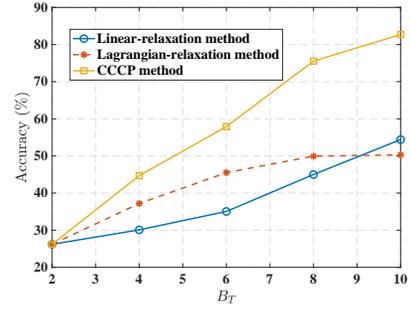}}\quad
  \subfigure[SKB at the receiver. ] {\includegraphics[width=.3\textwidth]{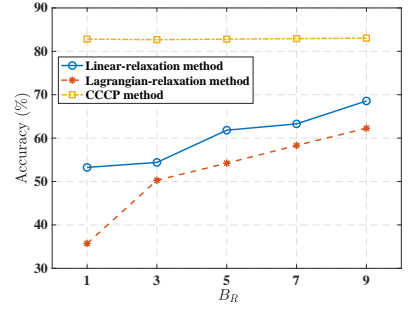}}\quad
  \caption{Effect of SKB on Transmission Performance.   }\label{fig:SKBtrans}
\end{figure}

\subsection{Tradeoff between the transmission latency and classification accuracy}
Table~\ref{tab:notation3} shows the tradeoff between the transmission latency and classification accuracy. The multi-level exatractor at the transmitter is assumed to be the same as that at the receiver. The SKB size at the transmitter is assumed to be full, i.e., $B_T = 10$, and that at the receiver is assumed to be $70\%$ of the total size of class prototype, i.e., $B_r = 7$. It can be observed that Level~1, Level~2, and Level~3 transmission achieve the same classification accuracy, while Level~2 incurs the least transmission latency.  This is because the dimension of intermediate feature at Level~2 is the smallest, and all the classification results are decided based on SKB at the receiver. Level~4 achieves the highest classification accuracy, while incurs larger latency than Level~2. This is because the SKB at the receiver is smaller than  that at the transmitter, and thus when more decisions made at the transmitter, i.e., Level~4, the classification accuracy would be higher. However, the dimension of semantic vector is higher than that of intermediate vector, and thus the transmission latency incurred by Level~4 is larger than that incurred by Level~2. Last but not the least, compared with the first three level transmission, CCCP can reduce the transmission latency requirement by $99.3\%$, $49.7\%$, $91.1\%$,  respectively, while achieving the same classification performance. Compared with Level~4 transmission, CCCP can reduce transmission latency requirement by $77.6\%$, without loss of classification performance by $1.2\%$. This is because CCCP can jointly consider both the transmission cost and the semantic loss.


\subsection{Effect of SKB at both the transmitter and receiver}

Fig.~\ref{fig:SKBtrans} (a) and Fig.~\ref{fig:SKBtrans} (b) illustrate the classification accuracy versus the size of SKB at the transmitter and that at the receiver, respectively. It can be seen that the classification accuracy increases with the size of SKB at the transmitter. This is because as the transmitter obtains more semantic knowledge, it can recognize the image class more accurately, and thus can only transmit the class index to the receiver. Thus, within a given transmission latency constraint, more images can be accurately recognized. In addition, the proposed CCCP method outperforms the other baselines, which indicates that CCCP can better utilize the knowledge at the transmitter. 

 Also, it can be seen that the classification accuracy increases with the size of SKB at the  receiver. This is because when the size of SKB at the receiver is relatively small, the transmitter has to deliver the exact semantic vector to the receiver for recognition, i.e., $3$rd-level semantic feature transmission. As the size of SKB at the receiver increases, the transmitter only has to deliver the estimated class index to the receiver, and the receiver searches its SKB to get the semantic feature, i.e., $4$th-level estimated class transmission. In addition, the proposed CCCP method outperforms the other baselines, which indicates that the CCCP can better utilize the knowledge at the receiver for remote zero-shot recognition.

\section{Conclusion}
This paper investigates a novel SKB-enabled E2E multi-level feature transmission framework for remote zero-shot recognition task. In particular, first, in order to serve the zero-shot learning task, an SKB-enabled multi-level feature extractor is established, which is not only lightweight, but also can facilitate communication overhead reduction. Then, the SKB-enabled multi-level feature transmission framework is constructed, and the corresponding semantic loss and transmission overhead are modeled.  The formulated semantic minimization problem, however, is a multi-choice knapsack problem, which is NP-hard and very challenging to solve. The CCCP algorithm is proposed to obtain an efficient sub-optimal solution.  Finally, numerical results are provided to verify the performance of the proposed designs. It is our hope that this paper can provide new insights on SKB construction for semantic communication, SKB-enabled semantic communication, etc. 

\bibliographystyle{IEEEtran}

\end{document}